\documentclass[twocolumn,superscriptaddress,prl,aps]{revtex4-1}
\usepackage{amsfonts}
\usepackage{amssymb}
\usepackage{amsmath}
\usepackage{graphicx}
\usepackage{epsfig}
\usepackage{color,xcolor}
\usepackage{hyperref}

\hypersetup{hypertex=true,colorlinks=true,linkcolor=blue,anchorcolor=blue,citecolor=blue}

\begin{document}

\title{Probing Bulk Band Topology from Time Boundary Effect in Synthetic
Dimension}
\author{Huisheng Xu}
\affiliation{School of Physics, Nankai University, Tianjin 300071, China}
\author{Zhaohui Dong}
\affiliation{State Key Laboratory of Advanced Optical Communication Systems and Networks,
School of Physics and Astronomy, Shanghai Jiao Tong University, Shanghai,
200240, China}
\author{Luqi Yuan}
\email{yuanluqi@sjtu.edu.cn}
\affiliation{State Key Laboratory of Advanced Optical Communication Systems and Networks,
School of Physics and Astronomy, Shanghai Jiao Tong University, Shanghai,
200240, China}
\author{Liang Jin}
\email{jinliang@nankai.edu.cn}
\affiliation{School of Physics, Nankai University, Tianjin 300071, China}

\begin{abstract}
An incident wave at a temporal interface, created by an abrupt change in
system parameters, generates time-refracted and time-reflected waves. We
find topological characteristics associated with the temporal interface that
separates distinct spatial topologies and report a novel bulk-boundary
correspondence for the temporal interface. The vanishing of either time
refraction or time reflection records a topological phase transition across
the temporal interface, and the difference of bulk band topology predicts
nontrivial braiding hidden in the time refraction and time reflection
coefficients. These findings, which are insensitive to spatial boundary
conditions and robust against disorder, are demonstrated in a synthetic
frequency lattice with rich topological phases engendered by long-range
couplings. Our work reveals the topological aspect of temporal
interface and paves the way for using the time boundary effect to probe
topological phase transitions and topological invariants.
\end{abstract}

\maketitle

\textit{Introduction.}---A temporal interface is the boundary in the time
domain where the system parameters undergo an abrupt change. The propagating
electromagnetic wave at the temporal interface separating two spatially
homogeneous media experiences time refraction and time reflection, governed
by a temporal analog of Snell's law based on the space-time duality of
Maxwell's equations \cite{Akhmanov,Kolner,ShuklaPS02}. Scattering at the
temporal interface resembles that at a spatial interface. Momentum (Energy)
is conserved before and after the temporal (spatial) interface due to
translational symmetry in space (time) \cite{AgrawalOL14,AgrawalPRL15}.
Thus, a change in the momentum of an incident wave at the spatial interface
becomes a change in the energy of the wave at the temporal interface.
However, an ideal temporal interface requires a rapid change in the system
within a duration much shorter than the wave dynamics, which is extremely
challenging to achieve in real materials. Recent experimental advancements
underscore the significant potential of fabricating temporal interfaces.
Although first observed for classical water wave \cite{FortNP16}, the
temporal interface for optical wave was later realized in an
epsilon-near-zero medium \cite{BoydNC20,HendryOptica21} and more recently
demonstrated for microwave \cite{AluNP23,PeroulisNC24} and acoustic wave
\cite{DaraioPRL24}.

Nowadays, the monolayer temporal interface opens up new opportunities for
manipulating the electromagnetic waves in the time domain \cite%
{TretyakovPRR24,SegevPRL24}. An immediate application is broadband frequency
translation \cite{BoydNC20,AluNP23}. The interference of time-refracted and
time-reflected waves from the multilayer temporal interfaces realizes
nonreciprocal \cite{AluPRL22}, reconfigurable \cite{PLuPNAS23}, and coherent
control \cite{GaliffiNP23,LYuan24}. Furthermore, periodically and abruptly
modulating the refractive index in time constructs photonic time crystals
\cite{MonticoneOME22,SegevOME24,Asadchy24}, which support nontrivial
momentum band topology \cite{MSegevOptica18}. Temporal topological interface
states within the momentum band gap are then identified as the time-domain
counterparts of spatial topological interface states \cite%
{SachaNJP19,JWDong23,YYang24}. This indicates a temporal analog of the
bulk-boundary correspondence \cite{Wen89,Hatsugai93}, which is a fundamental
principle of topological phases of matter: a topological invariant
characterizes the bulk band topologies of topological phases and predicts
interface states at the boundary separating regions with distinct band
topologies \cite{HasanRMP10,DasRMP16,TOzawaRMP19,HPriceJPP22}. Although
spatial (temporal) topological interface states arising from distinct band
topologies in the spatial (temporal) domain have been reported, to the best
of our knowledge, the temporal interface of a topological system in the
spatial domain remains unexplored. Recently, such a temporal interface in
the Su-Schrieffer-Heeger lattice \cite{SSH} was realized experimentally with
ultracold atoms, demonstrating the temporal analog of Snell's law \cite%
{BYanNatPhoton24,HannafordNP24}. Now the question is what topological
characteristics are associated with the temporal interface that separates
distinct spatial topologies.

Here, we showcase a novel bulk-boundary correspondence for such temporal
interfaces. The time refraction or time reflection vanishes at the momentum
of the degenerate point, associated with the topological phase transition
across the temporal interface. Furthermore, the time refraction and time
reflection coefficients braid in the Brillouin zone, and their linking
number equals the difference in the winding numbers across the temporal
interface. Consequently, the time boundary effect can probe topological
phase transition and band topology, fundamentally different from directly
measuring the bulk topological invariants \cite%
{IBlochNP13,YDChongPRX15,MassignanNC17,GCGuoPRL18,XMJinPRL19,EckardtPRL19,XJinPRL21,LQYuanLight23,St-JeanPRL24}
and the edge states \cite{HafeziNP13,HafeziNP16,HBSunPRL22,HSXuPRA22}. Our
findings are demonstrated in a synthetic frequency lattice, where frequency
space is analogous to spatial space \cite%
{LYuanOL16,TOzawaPRA16,ZYangPRX20,SFanSciece20,SFanLight20,YuPRL24,OzawaCP24}%
. We highlight the robustness of topological characteristics hidden in the
time boundary effect. This work creates new opportunities for the time
boundary effect in topological photonics.

\begin{figure}[tb]
\includegraphics[bb=0 0 265 150, width=8.8 cm, clip]{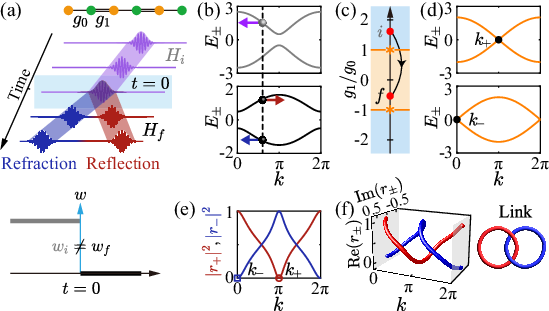}
\caption{Probing bulk band topology from the time boundary effect in the Su-Schrieffer-Heeger model.
The temporal interface is at $t=0$. The couplings before (after) the temporal interface are $g_0=1$, $g_1=1.5$ ($g_0=1$, $g_1=-0.5$). (a) Time refraction and reflection at the temporal interface.
(b) Energy bands before (gray) and after (black) the temporal interface. (c) Phase diagram. Topological phase transition occurs at $g_1=\pm g_0$ in orange. (d) Energy bands at $g_1=g_0$ (upper panel) and $g_1=-g_0$ (lower panel). (e) Zeros in $r_{+}(k)$ and $r_{-}(k)$ at the band gap closing degenerate points $k_{+}$ and $k_{-}$. (f) Braiding of $r_{+}(k)$ and $r_{-}(k)$.}
\label{fig1}
\end{figure}

\textit{Topological effects for temporal interface.}---The temporal
interface of a two-band photonic lattice in the spatial domain splits the
incoming light wave into two components [Fig.~\ref{fig1}(a)]: a
time-refracted wave that propagates in the same direction, and a
time-reflected wave that propagates in the opposite direction \cite%
{SFanPRR23}. The energy bands and eigenstates of the Bloch
Hamiltonian $H(k)$ of the two-band photonic lattice are denoted by $E_{\pm
}(k)$ and $\left\vert \psi _{\pm }\left( k\right) \right\rangle $, where $k$ is the momentum, and the subscript $+$ ($-$) represents the
upper (lower) band. Furthermore, we use additional subscripts $i$
and $f$ in $H(k)$, $E_{\pm }(k)$, and $\left\vert \psi _{\pm }\left(
k\right) \right\rangle $ to denote the Bloch Hamiltonians, energy bands, and
eigenstates before and after the temporal interface, respectively. Without
loss of generality, we choose the upper band $\left\vert \psi _{i,+}\left(
k\right) \right\rangle $ as the initial excitation arriving at the temporal
interface. Then, the time evolution is $|\Psi (t)\rangle
=r_{+}(k)e^{-iE_{f,+}(k)t}\left\vert \psi _{f,+}\left( k\right)
\right\rangle +r_{-}(k)e^{-iE_{f,-}(k)t}\left\vert \psi _{f,-}\left(
k\right) \right\rangle $. The projections $r_{\pm}(k)$ yield the time
refraction and reflection coefficients
\begin{equation}
r_{+}(k)=\left\langle \psi _{f,+}(k)\right\vert \left. \psi
_{i,+}(k)\right\rangle ,r_{-}(k)=\left\langle \psi _{f,-}(k)\right\vert
\left. \psi _{i,+}(k)\right\rangle .
\end{equation}%
The total probability is unity $|r_{+}(k)|^{2}+|r_{-}(k)|^{2}=1$ \cite{BYanNatPhoton24,SFanPRR23}. The velocities of time
refraction and reflection are predicted by $dE_{f,\pm }(k)/dk$ [Fig.~\ref%
{fig1}(b)].

When the Bloch Hamiltonians $H_{i}(k)$ and $H_{f}(k)$ share identical
eigenstates, the orthogonality between their eigenstates results in the
vanishing of time refraction or time reflection. This occurs at the momentum
of the band gap closing degenerate point associated with the topological
phase transition across the temporal interface (Supplemental Material A \cite%
{SI}). Notably, the topological phases chosen before and after the temporal
interface are represented by two red dots in the phase diagram [Fig.~\ref%
{fig1}(c)]. The straight line connecting the two red dots intersects the
topological phase boundaries at two orange asterisks. At the orange asterisk
inside (outside) the two red dots, the band gap closes at the momentum $%
k_{+} $ ($k_{-}$) [Fig.~\ref{fig1}(d)], where $r_{+}(k)$ [$r_{-}(k)$]
reaches zero [Fig.~\ref{fig1}(e)]. Thus, the zeros in the time refraction
and time reflection record the topological phase transitions. Furthermore,
when the topological phases before and after the temporal interface are
adjacent in the phase diagram, the minimal number of zeros in $r_{+}(k)$ and
$r_{-}(k)$ determines the minimal number of degenerate points encountered
during their topological phase transition. These degenerate points are
associated with changes in the band topology. This allows the detection of
topological invariant in the phase diagram tomograph by counting zeros in $%
r_{+}(k)$ and $r_{-}(k)$.

The synthetic frequency lattice with chiral symmetry enables the
visualization of topological effects emerging from the temporal interface.
The associated Bloch Hamiltonian has an off-diagonal form
\begin{equation}
H\left( k\right) =\left(
\begin{array}{cc}
0 & \mathcal{G}^{\dagger }\left( k\right) \\
\mathcal{G}\left( k\right) & 0%
\end{array}%
\right) .  \label{Hk}
\end{equation}%
The eigenenergies are $E_{\pm }(k)=\pm |\mathcal{G}(k)|$, and the
eigenstates are $\left\vert \psi _{\pm }(k)\right\rangle =(1,\pm e^{i\varphi
(k)})^{T}/\sqrt{2}$, where $\varphi (k)=\arg [\mathcal{G}(k)]$.~The
topological phase before (after) the temporal interface is characterized by
the winding number $w_{i(f)}=(2\pi )^{-1}{\int_{0}^{2\pi }\partial
_{k}\varphi _{i(f)}(k)dk}$, which is the number of times the complex $%
\mathcal{G}_{i(f)}(k)$ winds around the origin \cite{Wen89}.~The time
refraction and reflection coefficients are $r_{\pm }\left( k\right) ={\{1\pm
e^{i[\varphi _{i}(k)-\varphi _{f}(k)]}\}/2}$. The complex $r_{+}(k)$ and $%
r_{-}(k)$ braid in the Re[$r_{\pm} (k)$]-Im[$r_{\pm} (k)$]-$k$
space [Fig.~\ref{fig1}(f)]. After mapping the entire period ${k}\in \left[
0,2\pi \right] $ of $r_{\pm }(k)$ onto a torus by taking $k$ as the toroidal
direction, $r_{+}(k)$ and $r_{-}(k)$ become two closed curves that form a
link. The linking number equals the difference in the winding numbers across
the temporal interface (Supplemental Material B \cite{SI})
\begin{equation}
\mathcal{L}=w_{i}-w_{f}.
\end{equation}%
Thus, the time boundary effect can extract the difference in winding numbers
across the temporal interface. Probing bulk band topology
through time boundary effect differs from directly measuring the Zak phase
\cite{ZakPRL89}.

The topological characteristics hidden in the time boundary effect reveal
the bulk-boundary correspondence for the temporal interface that separates
distinct spatial topologies, including (i) the vanishing of time refraction 
or time reflection at the momentum of the band gap closing degenerate point,
and (ii) the braiding of time refraction and reflection arising from
distinct spatial topologies. These features uncover the topological phase
transition and the winding numbers across the temporal interface.

\textit{Rich topological phases.---}We propose a concrete synthetic
frequency lattice that supports rich topological phases with high winding
numbers to demonstrate the bulk-boundary correspondence for the temporal
interface. The multiplicity of band topologies is enriched by incorporating
long-range couplings \cite%
{YDChongPRL18,MSegevNature19,ABKNP20,CerjanPRL22,KhanikaevPRB22,BuljanNP23,ZYLiuPRAppl23}%
, which are typically weak or negligible due to the exponential decay of
evanescent fields \cite{GJTLPR22,Rho20}. The synthetic frequency dimension
offers a feasible platform for studying topological physics with long-range
couplings \cite{McMahonNP23,HPriceNRP19,SFanNature21,LYuanLight24}.

\begin{figure}[tb]
\centering\includegraphics[bb=0 0 260 105, width=8.8 cm, clip]{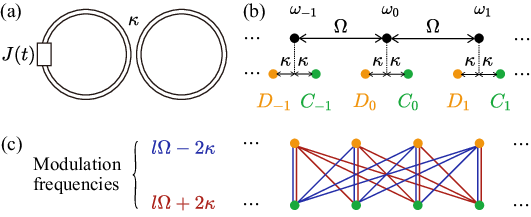}
\caption{(a) Coupled ring resonators. (b) The frequencies of symmetric  (green dots) and anti-symmetric (orange dots) supermodes.
(c) Schematic of the effective couplings created by the modulation frequencies in the sinusoidal modulation $J(t)$. The associated $H(k)$ has the form of Eq.~(\ref{Hk}).}
\label{fig2}
\end{figure}

We consider two identical ring resonators coupled with strength $\kappa $
[Fig.~\ref{fig2}(a)]. Each ring resonator supports a set of equally-spaced
resonant frequency modes $\omega _{n}=\omega _{0}+n\Omega $ ($n\in \mathbb{Z}
$) that differ by the free spectral range $\Omega $ \cite{LYuanOL16}, where $%
\omega _{0}$\ is the central frequency. The coupling $\kappa $ results in
symmetric supermodes $\omega _{n}+\kappa $ and antisymmetric supermodes $%
\omega _{n}-\kappa $ [Fig.~\ref{fig2}(b)] \cite%
{LQYuanLight23,St-JeanPRL24,SFanLight20}. An electro-optic phase modulator
is placed inside one of the two rings with an external modulation $J(t)$.
The sinusoidal modulation of the refractive index in the ring resonator can
establish connectivity between the supermodes \cite{LQYuanLight23}. The
spacing of the connectivity depends on the modulation frequency, while the
coupling strength is determined by the modulation amplitude, enabling the
realization of long-range couplings by choosing the modulation frequency as
a multiple of free spectral range \cite{YuanPRB18,McMahonNP23,SFanLight23}.
The synthetic frequency lattice is a ladder as illustrated in Fig.~\ref{fig2}%
(c). The modulation frequencies $l\Omega -2\kappa $ (blue) and $l\Omega
+2\kappa $ (red) lead to the inter-ladder-leg couplings, which appear as
off-diagonal terms in $H(k)$ (Supplemental Material C \cite{SI}). The
indices $l=0$, $\left\vert l\right\vert =1$, and $\left\vert l\right\vert >1$
are integers, and the associated modulations generate vertical
coupling, nearest-neighbor coupling, and long-range coupling, respectively
\cite{St-JeanPRL24}.

\begin{figure}[t]
\centering\includegraphics[bb=0 0 250 285, width=8.8 cm, clip]{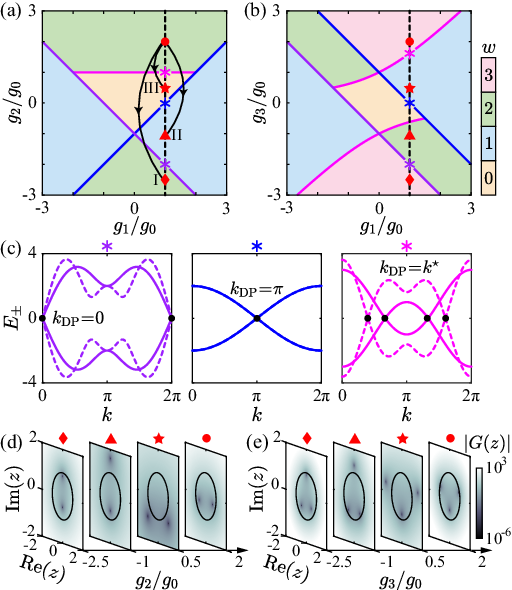}
\caption{Phase diagrams for the synthetic frequency lattice with long-range
coupling: (a) $g_2$ and (b) $g_3$. (c) Energy bands at the asterisks on
different phase boundaries. The solid and dashed lines are for (a) and (b), respectively. $k^{\star}=\arccos[-g_{1}/(2g_{2})]$ for (a) and $k^{\star}=\arccos[g_{0}/(2g_{3})]$ for (b). (d) and (e) $|G(z)|$ for the red diamond,
triangle, star, and dot in (a) and (b). The black ellipse represents the unit circle.
} \label{fig3}
\end{figure}

We focus on a polychromatic modulation $J(t)=\sum_{l}4g_{l}\cos \left[
\left( l\Omega -2\kappa \right) t+\phi _{l}\right] $. The corresponding
Bloch Hamiltonian $H(k)$ is Eq.~(\ref{Hk}) with $\mathcal{G}%
(k)=\sum_{l}g_{l}e^{i\phi _{l}}e^{ikl}$, preserving chiral symmetry. In the
ten-fold way classification of topological phases \cite{AZ1997}, $H(k)$
belongs to class BDI for $\phi _{l}=m\pi $ ($m\in \mathbb{Z}$), preserving
additional time-reversal symmetry and particle-hole symmetry; and $H(k)$
belongs to class AIII for $\phi _{l}\neq m\pi $ ($m\in \mathbb{Z}$), where
the phase factor $\phi _{l}$ creates a gauge field and affects the band
topology. Both classes feature a $\mathbb{Z}$ topological invariant \cite%
{RyuRMP16}. The winding number $w$ characterizes the band topology.
According to Cauchy's argument principle, $w$ equals the number of zeros
minus the order of pole from $G(z)=\sum_{l}g_{l}e^{i\phi _{l}}z^{l}$ in the
complex plane $z$, within the unit circle (i.e., $z=e^{ik} $). We discuss
the scenario $\phi _{l}=0$, and other scenarios can be treated similarly.

\begin{figure}[t]
\centering\includegraphics[bb=0 0 285 220, width=8.8 cm, clip]{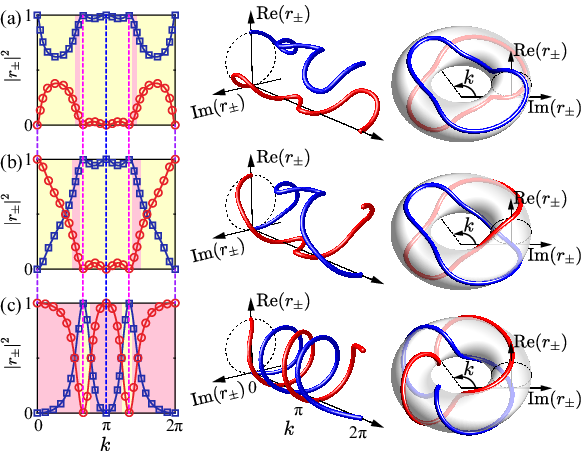}
\caption{The plots of $r_{+}(k)$ (red curve) and $r_{-}(k)$ (blue curve). As illustrated in Fig.~\protect\ref{fig3}(a), the topological phases across the time boundary lie along the dashed black line $g_{1}=g_{0}$. The topological phase before the time boundary at
$g_{2}=2g_{0}$ (red dot) has $w_{i}=2$, while the topological phases after the time boundary, at (a) $g_{2}=-5g_{0}/2$ (red diamond), (b) $g_{2}=-g_{0}$ (red triangle), and (c) $g_{2}=g_{0}/2$ (red star) for case I, II, and III, have $w_{f}=2$,  $w_{f}=1$, and $w_{f}=0$, respectively. Here, $r_{+}(k)$ represents time refraction (reflection), and $r_{-}(k)$ represents time reflection (refraction) within the pink (yellow) region (Supplemental Material E \cite{SI}). The markers correspond to numerical simulations~\protect\cite{GW}. The toroidal direction of the torus is the one-dimensional excitation momentum $k$. The cross section of the torus is $r_{\pm}(k)$.}
\label{fig4}
\end{figure}

A bichromatic modulation $J(t)$ with $g_{0},g_{1}\neq 0$ constructs the
Su-Schrieffer-Heeger model \cite{LQYuanLight23}, which supports a topologically trivial phase ($w=0$) and a topologically nontrivial phase ($w=1$). The long-range coupling $g_{l}$
with index $\left\vert l\right\vert >1$ creates topologically
nontrivial phases with high winding numbers \cite{St-JeanPRL24}. The phase
diagrams for the trichromatic modulation $J(t)$ with $g_{0},g_{1},g_{2}\neq
0 $ and $g_{0},g_{1},g_{3}\neq 0 $ are displayed in Figs.~\ref{fig3}(a) and %
\ref{fig3}(b) (Supplemental Material D \cite{SI}). At topological phase
transitions [colored asterisks in Figs.~\ref{fig3}(a) and \ref{fig3}(b)],
the band gap closes at the degenerate points of the energy bands [black dots
in Fig.~\ref{fig3}(c)], which are associated with a change in the band
topology. The different topologies in Figs.~\ref{fig3}(a) and~\ref{fig3}(b),
obtained from $|G(z)|$, are shown in Figs.~\ref{fig3}(d) and~\ref{fig3}(e),
where the pole is absent and the number of zeros within the unit circle
indicates the winding number of each phase.

\textit{Topological time boundary}.---We demonstrate the detection of bulk
band topology from the time boundary effect. Topological phases before and
after the time boundary, for the examples shown in Fig.~\ref{fig4}, are selected on the dashed black line in the phase diagram of
Fig.~\ref{fig3}(a). This dashed black line intersects the topological phase
boundaries at the colored asterisks. The band gap of $H(k)$ at the purple,
blue, and magenta asterisks closes at the degenerate points with the momenta
$k_{\mathrm{DP}}=0$, $k_{\mathrm{DP}}=\pi $, and ${k_{\mathrm{DP}}=2\pi /3,}%
4\pi /3$, respectively, as shown in Fig.~\ref{fig3}(c). In Fig.~\ref{fig4} (left panel), a remarkable feature is the vanishing of time
refraction or time reflection at these momenta. The dashed lines indicate
these zeros, using the colors of the corresponding topological phase
boundaries. Notably, these zeros, considered collectively, are
identical in all three cases, but their distributions in $r_{+}(k)$ and $%
r_{-}(k)$ are different.

Case I experiences all the purple, blue, and magenta topological phase
transitions [Fig.~\ref{fig3}(a)]. As a result, all the zeros at {$k_{\mathrm{%
DP}}=0$, $k_{\mathrm{DP}}=\pi $, and ${k_{\mathrm{DP}}=2\pi /3,}4\pi /3$}
appear in $r_{+}(k)$ [Fig.~\ref{fig4}(a)]. Similarly, case II experiences
the blue and magenta topological phase transitions [Fig.~\ref{fig3}(a)].
Consequently, the zeros at {$k_{\mathrm{DP}}=\pi $ and ${k_{\mathrm{DP}%
}=2\pi /3,}4\pi /3$} appear in $r_{+}(k)$, while the zero at $k_{\mathrm{DP}%
}=0$ appears in $r_{-}(k)$ [Fig.~\ref{fig4}(b)]. Finally, case III
experiences the magenta topological phase transition [Fig.~\ref{fig3}(a)].
Thus, the zeros at $k_{\mathrm{DP}}=2\pi /3,4\pi /3$ appear in $r_{+}(k)$,
and the zeros at {$k_{\mathrm{DP}}=0$ and $k_{\mathrm{DP}}=\pi $} appear in $%
r_{-}(k)$ [Fig.~\ref{fig4}(c)]. Furthermore, the interchange of zeros
between $r_{+}(k)$ and $r_{-}(k)$ signifies a topological phase transition,
and the associated momentum pinpoints the location of degenerate point. In
comparison with the zeros in Figs.~\ref{fig4}(a) and~\ref{fig4}(b), the zero
at $k_{\mathrm{DP}}=0$ in $r_{+}(k)$ switches to $r_{-}(k)$. This indicates
a topological phase transition, with the band gap closing at $k_{\mathrm{DP}%
}=0$, and $w$ changes by $1$. In comparison with the zeros in Figs.~\ref%
{fig4}(b) and~\ref{fig4}(c), the zero at $k_{\mathrm{DP}}=\pi $ in $r_{+}(k)$
switches to $r_{-}(k)$. This indicates another topological phase transition,
with the band gap closing at $k_{\mathrm{DP}}=\pi $, and $w$ changes by $1$.

In Fig.~\ref{fig4}, the braidings (middle panel) of time
refraction and reflection form links (right panel) when mapped onto a torus.
Figure~\ref{fig4}(a) shows case I, where $r_{+}(k)$ and $r_{-}(k)$ do not
braid, forming an unlink with $\mathcal{L}=0$. Figure~\ref{fig4}%
(b) shows case II, where $r_{+}(k)$ and $r_{-}(k)$ braid once, forming a
Hopf link with $\mathcal{L}=1$. Figure~\ref{fig4}(c) shows case III, where $%
r_{+}(k)$ and $r_{-}(k)$ braid twice, forming a Solomon link with $\mathcal{L%
}=2$. These links are topologically inequivalent and cannot be continuously
deformed from one to another without untying. The linking number
$\mathcal{L}$ of $r_{+}(k)$ and $r_{-}(k)$ accurately reflects the variation in the winding number across the time boundary $%
w_{i}-w_{f}$ \cite{Sign}, as verified from $w_{i}=2$ and $%
w_{f}=2,1,0$ for cases I, II, III. The formation of links, ensured by
different winding numbers across the time boundary, gives rise to $|r_{+}(k_{\mathrm{DQPT}}|^{2}=|r_{-}(k_{\mathrm{DQPT}})|^{2}=1/2$ for a dynamical quantum phase transition \cite%
{BalatskyPRL16}. Thus, the existence of $k_{\mathrm{DQPT}}$ is topologically
protected, and the least number of $k_{\mathrm{DQPT}}$ in the entire period $%
k\in \lbrack 0,2\pi ]$ is predicted by the link crossing number $2\mathcal{L}
$, i.e., the minimal number of crossings that occur in any projection of a
link. We highlight that the zeros and braidings associated with the time
refraction and time reflection are robust to disorder (Supplemental Material
F \cite{SI}). Thus, the time boundary effect is a versatile tool for phase
diagram tomography.

\textit{Conclusion and Discussion}.---The temporal degree of
freedom heralds a new era in the control and manipulation of light~\cite%
{TretyakovOME24}. The time boundary effect encodes information about bulk
band topologies both before and after the temporal interface. We uncover a
novel bulk-boundary correspondence for the temporal interface that separates
distinct spatial topologies and demonstrate the use of the time boundary
effect to probe bulk band topology. The vanishing of either time refraction or time
reflection signifies a topological phase transition across the temporal
interface, and the braiding of time refraction and time reflection
identifies different winding numbers across the temporal interface. Probing
topology using the time boundary effect is insensitive to the boundary
conditions of the synthetic frequency lattice, as the dynamics are confined
within a finite-size region. Furthermore, the vanishing of time refraction
or time reflection also occurs at the topological phase transition across
the temporal interface in high-dimensional topological phases~\cite{SI}.
Consequently, the time boundary effect in two-dimensional Chern insulators
can probe the topological phase transitions and the Chern numbers~\cite%
{GoldmanNP15,PeschelNP17,WeitenbergNC19,LeykamNP21,LeykamPRL21,XJLiuPRL23}.
Our findings pave the way for future investigations into nonequilibrium \cite%
{LQYuanLight21,HatsugaiPRL21}, non-Abelian \cite%
{GMaNP22,CTChanScience24,SFanNature24}, nonlinear \cite%
{YDChongPRL16,Yuri20,RechtsmanPRL21,Zilberberg24}, and non-Hermitian \cite%
{QGongNP24,SzameitNM24} topologies in spatiotemporal metamaterials \cite%
{SegevOptica22,PengPRL22,Szameit24}.

This work was supported by National Natural Science Foundation of China
(Grants No.~12222504, No.~12122407, No.~124B2079, and No.~12475021), and
National Key Research and Development Program of China (2023YFA1407200).


\begin{thebibliography}{999}
\bibitem{GW} The initial state is a Gaussian wave packet that
excites the eigenstate of the upper band. At the moment $t=0$, the
amplitudes of antisymmetric ($\psi _{D_{m}}$) and symmetric ($\psi _{C_{m}}$%
) supermodes in the $m$th unit cell are in the form of $\left\vert \Psi
_{m}(0)\right\rangle \equiv \left( \psi _{D_{m}},\psi _{C_{m}}\right)
^{T}=(e^{-(m-m_{c})^{2}/(2\sigma ^{2})}e^{ikm}/\sqrt[4]{\pi \sigma ^{2}}%
)\left\vert \psi _{i,+}(k)\right\rangle $, where $m_{c}=0$ governs the
center of the wave packet, and $\sigma=20 $ controls the width of the wave
packet.
\bibitem{Sign} The number of topological interface states formed at the
spatial interface is the change in the winding number across the spatial
interface, $|w_i-w_f|$ \cite{LLuNP16,ABKNP17,QGongAOM17}.
\end{thebibliography}
\end{document}